\let\ifarxiv=\iftrue     
\pdfoutput=1

\ifarxiv

\documentclass[12pt,a4paper]{article}
\usepackage[a4paper,text={450pt,650pt},centering]{geometry}

\fi

\ifarxiv\else

\documentclass[11pt,a4paper]{article}
\usepackage{mathptmx}
\usepackage[a4paper,text={130mm,198mm}]{geometry}

\fi

\usepackage{amsmath,amssymb}
\usepackage[bookmarks=true,hyperfigures=true]{hyperref}
\usepackage{graphicx}
\usepackage[nosort]{cite}
\usepackage[bulletsep]{collref}

\let\oldbfseries=\bfseries
\let\oldmdseries=\mdseries
\let\oldnormalfont=\normalfont
\renewcommand{\bfseries}{\oldbfseries\boldmath}
\renewcommand{\mdseries}{\oldmdseries\unboldmath}
\renewcommand{\normalfont}{\oldnormalfont\unboldmath}

\allowdisplaybreaks[3]

\numberwithin{equation}{section}

\usepackage[font=small,labelfont=bf,width=0.85\textwidth]{caption}

\providecommand{\hypersetup}[1]{}

\hypersetup{plainpages=false}
\hypersetup{pdfpagemode=UseNone}
\hypersetup{bookmarksnumbered=true}
\hypersetup{pdfstartview=FitH}
\hypersetup{colorlinks=false}
\hypersetup{citebordercolor={.5 1 .5}}
\hypersetup{urlbordercolor={.5 1 1}}
\hypersetup{linkbordercolor={1 .7 .7}}

\providecommand{\arxivref}[2]{\href{http://arxiv.org/abs/#1}{#2}}
\providecommand{\doiref}[2]{\href{http://dx.doi.org/#1}{#2}}

\providecommand{\href}[2]{#2}
\providecommand{\arxivlink}[1]{\href{http://arxiv.org/abs/#1}{arxiv:#1}}

\newcommand{\pint}{\makebox[0pt][l]{\hspace{2.4pt}$-$}\int}

\begin{document}


\thispagestyle{empty}
\phantomsection
\addcontentsline{toc}{section}{Title}

\begin{flushright}\footnotesize%
\texttt{UUITP-40/10},
\texttt{\arxivlink{1012.3993}}\\
overview article: \texttt{\arxivlink{1012.3982}}%
\vspace{1em}%
\end{flushright}

\begingroup\parindent0pt
\begingroup\bfseries\ifarxiv\Large\else\LARGE\fi
\hypersetup{pdftitle={Review of AdS/CFT Integrability, Chapter III.4: Twist states and the cusp anomalous dimension}}%
Review of AdS/CFT Integrability, Chapter III.4:\\
Twist states and the cusp anomalous dimension
\par\endgroup
\vspace{1.5em}
\begingroup\ifarxiv\scshape\else\large\fi%
\hypersetup{pdfauthor={Lisa Freyhult}}%
Lisa Freyhult
\par\endgroup
\vspace{1em}
\begingroup\itshape
Department of Physics and Astronomy, Division for Theoretical Physics, Uppsala University, Box 516, SE-751 20 Uppsala, Sweden
\par\endgroup
\vspace{1em}
\begingroup\ttfamily
lisa.freyhult@physics.uu.se
\par\endgroup
\vspace{1.0em}
\endgroup

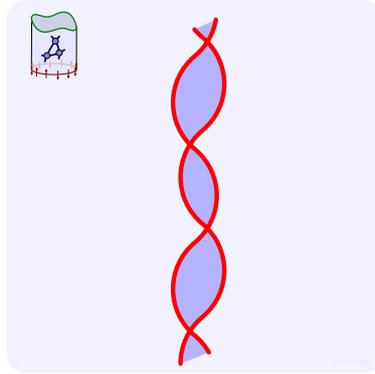
\begin{center}
\includegraphics[width=5cm]{TitleIII4.mps}
\vspace{1.0em}
\end{center}

\paragraph{Abstract:}
We review the computation of the anomalous dimension of twist operators in the planar limit of  $\mathcal{N}=4$ SYM using the asymptotic Bethe ansatz and demonstrate how this quantity is obtained at weak, strong and intermediate values of the coupling constant. The anomalous dimension of twist operators in the limit of large Lorentz spin played a major role in the construction as well as in many tests of the asymptotic Bethe equations, this aspect of the story is emphasised.

\ifarxiv\else
\fi
\hypersetup{pdfsubject={MSC (2010): 81T13, 81T60, 81T30, 83E30}}%

\ifarxiv\else
\fi
\hypersetup{pdfkeywords={twist operators, anomalous dimension, AdS/CFT, integrable models}}%

\newpage

\section{Introduction}
One of the best investigated quantities in the subject of integrability in the context of the AdS/CFT correspondence is the anomalous dimension of twist operators\footnote{This name might be slightly misleading. The twist of an operator is defined as its scaling dimension minus its Lorentz spin. Hence one could refer to any local operator with a definite value of the twist as a twist operator. We will make it explicit below which operators we are considering. In the literature these kind of operators are also referred to as Wilson operators.}.
The reasons for this are numerous and range from the interest in these operators due to the role they play in deep inelastic scattering in QCD to the fact that their anomalous dimension provides a quantity ideal for studying the AdS/CFT correspondence, a quantity that interpolate between the weak and strong coupling regimes of the theory.

The construction of the asymptotic Bethe equations that determine the spectrum of planar $\mathcal{N}=4$ supersymmetric Yang-Mills theory was based on and inspired by a number of conjectures. The two major ones are the complete, all-loop, integrability of the model and the AdS/CFT correspondence itself. 
Given the conjectures at the basis of the asymptotic construction it is absolutely necessary to put it to as many tests as possible and for this suitable quantities, amenable for studies, are needed. Twist operators constitute such quantities and played a major role in the developments  of the asymptotic system and were also important for understanding its inadequacies. 
It is now beyond any doubt that the asymptotic equations do not provide the final answer to the spectral problem in the planar limit. A system taking finite size effects into account is being constructed and also for tests of this system twist operators have played and can be expected to play an important role, see \cite{chapLuescher,chapTrans,chapTBA} for reviews.

The operators in question appear in a wide range of contexts, see also the review \cite{chapQCD}. Their anomalous dimensions can be computed from considerations of Wilson loops, gluon scattering amplitudes and they, as mentioned, play an important role in QCD and deep inelastic scattering. Naturally this is a great advantage since this allows for many cross-checks. Furthermore, providing all loop expressions for the scaling dimensions of the operators is not only important to AdS/CFT and the studies of integrability but also for making progress in these other areas.
 
The simplest representatives of twist operators in $\mathcal{N}=4$ SYM are operators in the $\mathfrak{sl}(2)$ subsector, constructed from complex scalar fields, $Z$, and covariant light-cone derivatives, $D$,
\begin{equation}\label{twistop}
O(x)=\mbox{Tr}(D^MZ^L)+\dots
\end{equation}
The abbreviation denotes all possible ways in which the derivatives are distributed on the scalar fields. The length or the twist of the operator is denoted by $L$ and $M$ is the number of covariant derivatives or the Lorentz spin.

The asymptotic Bethe equations were derived by assuming all loop integrability and imposing $PSU(2,2|4)$ symmetry on the internal S-matrix of the theory, see \cite{chapABA,chapSMat,chapSProp} for a review. This fixes the S-matrix and the Bethe equations up to an overall phase  \cite{Staudacher:2004tk,Beisert:2005fw}. In analogy with relativistic models the phase obeys crossing symmetry \cite{Janik:2006dc,Gomez:2006va,Plefka:2006ze}, this symmetry however does not completely constrain the phase. The condition of crossing symmetry has to be supplemented by information on the analytical properties of the phase. A proposal for the dressing phase was made based upon the structural properties of the anomalous dimensions of twist operators in the limit of large spin, M \cite{Eden:2006rx,Beisert:2006ez}. An important clue was provided by the impressive direct field theory computation of the same object to four loops \cite{Anastasiou:2003kj,Bern:2005iz,Bern:2006ew} which allowed to test the proposed Bethe equations and the dressing phase. Later it was demonstrated that the conjectured phase also constitutes a minimal solution to the crossing equations \cite{Volin:2009uv}. 

In the limit of large spin the anomalous dimension of the operators exhibits logarithmic, Sudakov, scaling \cite{Collins:1989bt, Belitsky:2006en}  and all the coupling constant dependence is collected in the so called scaling function. 
The scaling function was also obtained from considerations of light like Wilson loops with a cusp \cite{Korchemsky:1985xj,Korchemsky:1988si,Korchemsky:1992xv}, both at weak and strong coupling \cite{Kruczenski:2002fb,Alday:2007hr}, and is therefore also termed the cusp anomalous dimension. 

Twist operators in the large spin limit have a universal behavior, their minimal anomalous dimension does not depend on the length of the operator \cite{Korchemsky:1988si,Korchemsky:1992xv,Belitsky:2003ys,Eden:2006rx}. It was therefore conjectured that this quantity was not affected by finite-size/wrapping effects (see \cite{chapHigher} for a review) and consequently the asymptotic Bethe equations can be used to compute it to any loop order. This is also in line with the identification with cusped Wilson loops which independently implies no wrapping corrections. This allows for the construction of an all loop integral equation \cite{Eden:2006rx,Beisert:2006ez} which, if solved, provides the anomalous dimension as a  function of the coupling constant, a quantity that smoothly interpolates between weak and strong coupling. Needless to say such a quantity is of great importance for a better understanding of the AdS/CFT correspondence.

The string duals of twist two operators are folded strings with spin $S$ on $AdS_3$. The string sigma model allows for a semiclassical expansion and the energy of the string states can be obtained to leading orders in the sigma model loop expansion, see the discussion in \cite{chapSpinning} and \cite{chapQString}. The spin $S$ is taken large and identified with $M$ and for the dual operators this provides a prediction of their scaling dimension at strong coupling. The agreement between the string energy \cite{Gubser:2002tv, Frolov:2002av, Roiban:2007jf, Roiban:2007dq,Beccaria:2010ry} and the anomalous dimensions of the dual operators found by solving the Bethe equations at strong coupling \cite{Kotikov:2006ts,Alday:2007qf, Kostov:2007kx,Beccaria:2007tk,Basso:2007wd,Casteill:2007ct,Kostov:2008ax,Basso:2009gh} provided a remarkable check of the equations as well as the AdS/CFT correspondence.

Here we will review the developments that led to an integral equation determining the all loop expressions for the scaling dimensions of twist operators in the limit of large Lorentz spin. We will then solve this equation at weak, strong and intermediate values of the coupling constant and relate the results obtained to other, completely independent, computations of the same object. We will further review the anomalous dimensions for finite values of the Lorentz spin as well as their subleading corrections in the large spin expansion and the structural properties of the weak and strong coupling expansions.

\section{The asymptotic Bethe ansatz for sl(2) twist operators}

The twist operators \eqref{twistop} belong to the $\mathfrak{sl}(2)$ subsector, a closed subsector of the theory. In the spin chain picture, using $\mbox{Tr}(Z^L)$ as the reference state the covariant derivatives are interpreted as excitations on this vacuum. Any number of excitations $M$ is hence allowed and this number may exceed the length of the operator. 
The Bethe equations which determine the asymptotic spectrum of anomalous dimensions in the $\mathfrak{sl}(2)$ sector are then written, in terms of the rapidities or Bethe roots $u_i$, as  \cite{chapABA,chapSMat,chapSProp}
\begin{equation}\label{BE}
\left(\frac{x^+_k}{x^-_k}\right)^L=\prod_{j=1}^M\frac{u_k-u_j-i}{u_k-u_j+i}\left(\frac{1-g^2/x_k^+x_j^-}{1-g^2/x_k^-x_j^+}\right)^2e^{2i\theta(u_k,u_j)}
\end{equation}
where $g^2=\frac{\lambda}{16\pi^2}$, $\lambda$ is the 't Hooft coupling, $u=x(u)+\tfrac{g^2}{x(u)}$ and $x^\pm(u)=x(u\pm i/2)$.
Solving the Bethe equations together with the cyclicity constraint $\prod_{j=1}^L\frac{x^+_j}{x^-_j}=1$ for the rapidities the anomalous dimension can be computed using
\begin{equation}\label{defAD}
\gamma(g)=2g^2E=2g^2\sum_j\left(\frac{i}{x^+_j}-\frac{i}{x^-_j}\right).
\end{equation}
Here $E$ denotes the energy of the corresponding spin state. The explicit form of the dressing phase, $\theta(u,v)$, can be found in \cite{Beisert:2006ez, chapSProp}.  At leading order in the weak coupling expansion \eqref{BE} and \eqref{defAD} give the spectrum of the non-compact $XXX_{-1/2}$ spin chain with nearest neighbor interactions.
In this sector all the Bethe roots are real.
In order to study the limit when $M\to\infty$ it is useful to first consider the leading order equations, determining the anomalous dimension to one-loop, on the form
\begin{equation}\label{logBEone}
-iL\log\frac{u_k+i/2}{u_k-i/2}=2\pi n_k-i\sum_{j\neq k}^M\log\frac{u_k-u_j-i}{u_k-u_j+i}.
\end{equation}
Here $n_k$ reflects the choice of branch of the logarithm. This choice specifies the state in the spectrum. For generic values of $L$ the states occupy a band \cite{Belitsky:2003ys,Belitsky:2006en,Belitsky:2008mg} and we will here make the choice to study the lowest state in the band by setting $n_k=\mbox{sign}(u_k)$ and consider configurations where the roots are distributed symmetrically around the origin\footnote{For twist $L\geq3$ it is possible to construct several operators with the same $L$ and $M$ with different anomalous dimensions. The anomalous dimensions can be labeled by an additional quantum number, $l$, as $\gamma_{L,M}^{(l)}$ and represents the eigenvalues of the mixing matrix associated with the dilatation operator. They can be ordered as $\gamma_{M,L}^{(0)}<\gamma_{M,L}^{(1)}<\dots$ and are smooth functions of $M$. We refer to $\gamma_{M,L}^{(0)}$ as the lowest state in the band and the others as excited states. The different states are distinguished in the Bethe ansatz by their different the mode numbers. See the references above for a detailed discussion.}. For $L=2$ there is only one state.

In the limit the roots 
accumulate on a smooth contour with endpoints $\pm b$ and the discrete roots in \eqref{logBEone} can be replaced by the continuum parameter $u$. Introducing the rescaled roots $\bar{u}=u/M$, the one-loop density $\rho_0(u)=\frac{1}{M}\sum_j\delta(u-u_j)$ and the rescaled density $\bar{\rho}_0(\bar{u})=\frac{1}{M}\sum_j\delta(\bar{u}-\tfrac{u_j}{M})$ with support on the interval $[-b,b]$ the leading  equation can be written as
\begin{equation}
0=2\pi\mbox{sign}(\bar{u})-2\pint_{-b}^b\frac{\bar{\rho}_0(\bar{u}')d\bar{u}'}{\bar{u}-\bar{u}'}.
\end{equation}
The integral with a bar in the above equation is used to denote the principal value integral.
A solution is found by standard methods using the inverse Hilbert transform \cite{Eden:2006rx}
\begin{equation}\label{rhoK}
\bar{\rho}_0(\bar{u})=\frac{1}{\pi}\log\frac{1+\sqrt{1-\bar{u}^2/b^2}}{1-\sqrt{1-\bar{u}^2/b^2}}.
\end{equation}
This result was initially obtained by an alternative method in \cite{Korchemsky:1995be}.
The normalization of the density,
\begin{equation}
\int_{-b}^b\bar{\rho}_0(\bar{u})d\bar{u}=1,
\end{equation}
 determines the endpoints to $\pm b=\pm1/2$. The one-loop anomalous dimension in the large $M$ limit is hence
 \begin{equation}\label{ADoneloop}
 \gamma_0=\frac{2g^2}{M}\int_{-1/2}^{1/2}\frac{\bar{\rho}_0(\bar{u})d\bar{u}}{\bar{u}^2+1/4M^2}=4g^2\log\frac{\sqrt{1+1/M^2}+1}{\sqrt{1+1/M^2}-1}=8g^2\log M+\mathcal{O}(M^0).
 \end{equation}
Having solved the one-loop problem we proceed to all loops.  Taking the logarithm of both sides of the equation \eqref{BE} and multiplying by $i$ we write
 \begin{eqnarray}
 \nonumber &&\hspace{-0.4cm}{2L\arctan(2u_k)+iL\log \frac{1+g^2/(x_k^-)^2}{1+g^2/(x_k^+)^2}=2\pi\tilde{n}_k-2\sum_{j=-M/2\atop j\neq k}^{M/2}\arctan(u_k-u_j)}\\
 &&\hspace{-0.4cm}+2i\sum_{j=-M/2}^{M/2}\log \frac{1-g^2/x_k^+x_j^-}{1-g^2/x_k^-x_j^+}-2\sum_{j=-M/2}^{M/2}\theta(u_k,u_j),\quad k=\pm1,\,\pm2,\,\dots,\pm \tfrac{M}{2}
 \end{eqnarray}
 where the shifted mode numbers are
 \begin{equation}
 \tilde{n}_k=\frac{L-3}{2}\mbox{sign}(k)+k=\frac{L-2}{2}\mbox{sign}(k)+k',\quad k'=\pm \tfrac{1}{2}, \pm \tfrac{3}{2},\dots,\pm\tfrac{M-1}{2}.
 \end{equation}
It is convenient to use shifted mode numbers, as opposed to the mode numbers we used at one-loop, as in the limit when $M$ is taken to infinity $x=k'/M$ becomes a continuous variable and the density can be obtained from $\rho(u)=\frac{dx}{du}$. Note however that this is just a trick used to get the equations on a convenient form, the state and mode numbers specified are identical to the one-loop case studied above. Differentiating we find the leading continuum equation
 \begin{eqnarray}\label{leadingeq}
 \nonumber0&=&2\pi\rho(u)-2\int_{-M/2}^{M/2}\frac{\rho(u')du'}{(u-u')^2+1}+2i\int_{-M/2}^{M/2}du'\rho(u')\frac{d}{du}\log\frac{1-g^2/x^+(u)x^-(u')}{1-g^2/x^-(u)x^+(u')}\\
 &-&2\int_{-M/2}^{M/2}\frac{d}{du}\theta(u,u')\rho(u')du'.
 \end{eqnarray}
 To this leading order all dependence on the twist $L$ is removed. Twist dependence will however enter as the first subleading corrections are included.
 By a numerical analysis of this equation \cite{Eden:2006rx} one finds that the density with the one-loop part subtracted is localised close to the origin. Splitting the density $\rho(u)=\rho_0(u)+\sigma(u)$ the limits of integration in all integrals containing $\sigma(u)$ can therefore be extended to $\pm\infty$ as the limit $M\to\infty$ is taken. The integrals containing $\rho_0(u)$ can then be explicitly evaluated using \eqref{rhoK}.
Finally rescaling $\sigma(u)\rightarrow -\frac{\gamma_0}{2M}\sigma(u)$ this leads to the following equation for the Fourier-Laplace transform\footnote{ For details on the Fourier-Laplace transforms see \cite{Eden:2006rx}.
}, $\hat{\sigma}(t)=e^{-t/2}\int_{-\infty}^\infty du\, e^{-itu}\sigma(u)$,
 \begin{equation}\label{BES}
 \hat{\sigma}(t)=\frac{t}{e^t-1}\left(K(2gt,0)-4g^2\int_0^\infty K(2gt,2gt')\hat{\sigma}(t')\right).
 \end{equation}
 The kernel in \eqref{BES} is given by
 \begin{eqnarray}
&&\nonumber K(t,t')=K_0(t,t')+K_1(t,t')+K_d(t,t')\\
&&\nonumber K_0(t,t')=\frac{tJ_1(t)J_0(t')-t'J_0(t)J_1(t')}{t^2-t'^2}=\frac{2}{tt'}\sum_{n=1}^\infty (2n-1)J_{2n-1}(t)J_{2n-1}(t')\\
&&\nonumber K_1(t,t')=\frac{t'J_1(t)J_0(t')-tJ_0(t)J_1(t')}{t^2-t'^2}=\frac{2}{tt'}\sum_{n=1}^\infty 2n\,J_{2n}(t)J_{2n}(t')\\
&&K_d(t,t')=8g^2\int_0^\infty dt'' K_1(t,2gt'')\frac{t''}{e^{t''}-1}K_0(2gt'',t'),
 \end{eqnarray}
where the kernel $K_d(t,t')$ has its origin in the dressing phase. From \eqref{defAD} we find that it is possible to rewrite the anomalous dimension in terms of the density at zero,
\begin{equation}\label{cuspAD}
\gamma(g)=16g^2\sigma(0)\log M=f(g)\log M.
\end{equation}
The function $f(g)$ is the so called scaling function. We note here that the higher loop density, $\sigma(u)$, also gives the one-loop anomalous dimension. Computing the anomalous dimension in this way, as opposed to directly from \eqref{defAD}, requires information about the density to one loop order higher than the desired anomalous dimension.

The leading integral equation for the density \eqref{BES} is independent of the twist of the operator or the length of the corresponding spin chain. The derivation above is based on the asymptotic Bethe ansatz which is expected to break down at order $g^{2L}$ due to wrapping effects. For twist two operators the asymptotic equations are, due to superconformal invariance, valid to $\mathcal{O}(g^6)$ \cite{Staudacher:2004tk,Beisert:2005fw} \footnote{States in the same supersymmetry multiplet have the same anomalous dimension. In the same multiplet as the operators \eqref{twistop} with $L=2$ there are also length four operators and therefore wrapping is delayed.} . The fact that the result is independent of the twist lead to the conjecture that the anomalous dimension is universal and that wrapping plays no role for these operators to the leading order in the large spin expansion \cite{Eden:2006rx}.

In the approach to the study of twist operators discussed above an operator constructed solely from scalar fields was identified with the ground state of the spin chain. The covariant derivatives were viewed as excitations on the chain and we had to solve the Bethe equations for a large number of excitations to be able to describe the states of interest. There is another possible description of the operators that is somewhat more natural in this context. It is possible to let the derivatives act as an effective ground state and instead view the scalar fields as excitations. This reduces the complexity of the problem as in this case the twist two operators can be described by considering the scattering of only 2 excitations, termed holes \cite{Belitsky:2006en,Freyhult:2007pz}. In this description the one-loop anomalous dimension for twist $L$ is written as
\begin{eqnarray}\label{ADholes}
\nonumber\frac{\gamma_0}{g^2}&=&4\gamma_E L+2\sum_{j=1}^L\left(\psi(1/2+iu_h^{(j)})+\psi(1/2-iu_h^{(j)})\right)\\
&+&2\int_{-\infty}^\infty\frac{dv}{\pi}i\frac{d^2}{dv^2}\left(\log \frac{\Gamma(1/2+iv)}{\Gamma(1/2-iv)}\right)\mbox{Im}\log\left[1+(-1)^\delta e^{iZ(v+i0)}\right]
\end{eqnarray}
where $\delta=L+M\,\,\,\mbox{mod}\,\,2$ and $Z(u)$ denotes the counting function which in turn is determined by a non-linear integral equation. This kind of equations are often referred to as Destri-de Vega equations or, simply, non-linear integral equations, NLIE, and have been discussed in numerous publications, see \cite{Feverati:2006tg,Feverati:2006hh,Fioravanti:2007un,Bombardelli:2007ed} for a pedagogical introduction and further references.
The counting function determines the rapidities of the holes as well as of the Bethe roots. In the limit of large $M$ one immediately finds two large holes with $u_h\to\pm M/\sqrt2$ and in addition it is possible to show that the non-linear term in \eqref{ADholes} go as $4\log 2+\mathcal{O}((\tfrac{\log M}{M})^2)$, see \cite{Freyhult:2007pz} for details. The leading anomalous dimension \eqref{ADoneloop} is therefore immediately obtained. This analysis extends to include all higher loops and it is also possible to treat general values of the twist. To leading order in the large $M$ expansion one finds this way the same integral equation as in \eqref{BES}. Making use of this method it is however straightforward to continue the expansion and include also subleading corrections, for general $L$ this allows for a the construction of an integral equation for terms proportional to $M^0$ \cite{Freyhult:2007pz,Bombardelli:2008ah,Freyhult:2009my} and all corrections of the form $1/(\log M)^k$, with $k$ any integer $\geq1$ \cite{Fioravanti:2009ei}. 
For general $L$ the anomalous dimension has the structure
\begin{equation}\label{structuretwistL}
\gamma_L(g,M)=f(g)\left(\log M+\gamma_E+(L-2)\log 2\right)+B_L(g)+\sum_{k=1}^\infty \frac{C_L^{(k)}(g)}{(\log M)^k}\dots
\end{equation}
For $L=2,3$ one can continue the expansion up to $\mathcal{O}((\tfrac{\log M}{M})^2)$ where the non-linear terms start contributing. For $L=2$ the result is
\begin{equation}\label{twist2subleading}
\gamma_{2}(g,M)=f(g)\left(\log M+\gamma_E+\frac{f(g)}{2}\frac{\log M+\gamma_E}{M}+\frac{1+B_2(g)}{2M}\right)+B_2(g)+\mathcal{O}((\tfrac{\log M}{M})^2).
\end{equation}
The advantage of this method is that it is completely straightforward. No splitting of densities or input from numerics, as was used to derive \eqref{BES} from \eqref{leadingeq}, was needed. On the other hand one here faces the difficulty of treating the non-linear term.

\section{Weak coupling expansion}

The evaluation of the scaling function at weak coupling constitutes an important test of the conjectured asymptotic all loop Bethe equations. This very same function appears in scattering amplitudes, in light-like Wilson loops with a cusp and is part of a direct Feynman-diagram evaluation of the corresponding QCD result. For this reason there are several ways to compute the quantity and predictions for what the Bethe equations, if correct, should give.

The integral equation \eqref{BES} is a Fredholm equation of the second type and easily expanded to many loop orders at weak coupling,
\begin{eqnarray}\label{weakcouplingscaling}
 f(g)&=&8g^2-\frac{8\pi^2}{3}g^4+\frac{88\pi^4}{45}g^6-16\left(\frac{73}{630}\pi^6+4\zeta(3)^2\right)g^8+\dots
\end{eqnarray}
This expansion, to any loop order, shows an important structure. Assigning {\sl the degree of transcendentality} $k$ to $\zeta(k)$ and $\pi^k$ we find that the $l$-loop term has degree of transcendentality $2l-2$ \footnote{The transcendentality of a product is given by the sum of the transcendentalities of the factors.}. This is a manifestation of the maximal transcendentality principle conjectured in \cite{Kotikov:2002ab}. This principle states that the $\mathcal{N}=4$ result can be extracted from the corresponding QCD result by removing all terms that are not of maximal transcendentality. Using this conjecture it was possible to extend the one-loop result, first computed in \cite{Georgi:1951sr,Gross:1973ju,Dolan:2000ut}, to two loops by extracting the scaling function from the QCD result \cite{Kotikov:2003fb}, obtained by a direct field theory calculation. Further the three loop result was obtained in \cite{Kotikov:2004er} using the computation of the QCD splitting functions in \cite{Moch:2004pa}.

 This prediction was confirmed using the fact that the scaling function determines the leading $1/\epsilon^2$ pole of the logarithm of gluon amplitudes computed using dimensional regularisation in $4-2\epsilon$ dimensions. The two and three loop planar four point amplitude in $\mathcal{N}=4$ SYM was computed in \cite{Anastasiou:2003kj,Bern:2005iz}. Through an impressive effort that computation was also extended to four loops \cite{Bern:2006ew}. 

The anomalous dimension of light-like Wilson loops with a cusp is identified with the anomalous dimension of twist operators and also provide the first orders in the weak coupling expansion  \cite{Korchemsky:1985xj,Korchemsky:1988si,Korchemsky:1992xv}.

Up to three loops the result \eqref{weakcouplingscaling} is not sensitive to the dressing phase, to be able to fix the dressing phase the knowledge of the four loop scaling function therefore played a major role \cite{Eden:2006rx,Bern:2006ew,Beisert:2006ez}. The scaling dimension, computed from the Bethe equations with the correct dressing phase, is in full agreement with the results of the completely independent calculations mentioned above, providing highly non-trivial evidence that the Bethe equations and the assumptions made to derive them are indeed correct.

 In fact the three-loop result extracted from the QCD field theory computation is exact in $M$ and naturally expressed in terms of harmonic sums
\begin{eqnarray}
\label{klov}
\lefteqn{\gamma(M) = 8g^2\, S_1-16g^4\Big( S_{3} + S_{-3}  -
2\,S_{-2,1} + 2\,S_1\,\big(S_{2} + S_{-2}\big) \Big)}
\nonumber \\
 &&-64g^6 \Big( 2\,S_{-3}\,S_2 -S_5 -
2\,S_{-2}\,S_3 - 3\,S_{-5}  +24\,S_{-2,1,1,1}\nonumber+ 
6\,\big(S_{-4,1} + S_{-3,2} + S_{-2,3}\big)
\\&&- 12\,\big(S_{-3,1,1} + S_{-2,1,2} + S_{-2,2,1}\big)\nonumber -
\big(S_2 + 2\,S_1^2\big) 
\big( 3 \,S_{-3} + S_3 - 2\, S_{-2,1}\big)\\
&&- S_1\,\big(8\,S_{-4} + S_{-2}^2 
4\,S_2\,S_{-2} +
2\,S_2^2 + 3\,S_4 - 12\, S_{-3,1} - 10\, S_{-2,2} 
+ 16\, S_{-2,1,1}\big)
\Big)\,,
\end{eqnarray}
where
\begin{eqnarray}
\label{harmonic}
S_a=~ \sum^M_{m=1} \frac{(\mbox{sign}(a))^m}{m^a},
\qquad
S_{a_1,a_2,a_3,\cdots}~=~ \sum^M_{m=1} \frac{(\mbox{sign}(a_1))^m}{m^{a_1}}\,
S_{a_2,a_3,\cdots}(m)\, . 
\end{eqnarray}
For properties of the generalised harmonic sums see for example \cite{Vermaseren:1998uu}.
The transcendentality principle for the expression exact in $M$ states that the degrees of the sums, $|a_1|+|a_2|+|a_3|+\dots$, should add up to $2l-1$ at $l$ loops.

The exact expressions \eqref{klov} can also be obtained from the Bethe ansatz. One way to obtain such expressions is to numerically solve the Bethe equations for different values of $M$ and then fit the obtained result to a linear combination of harmonic sums and products of harmonic sums that obeys the transcendentality principle \cite{Beccaria:2007cn,Kotikov:2007cy,Beccaria:2007bb, Beccaria:2009eq, Lukowski:2009ce}. While there are additional properties that allows to restrict the number of terms in such a ansatz, see \cite{Beccaria:2007cn,Kotikov:2007cy,Beccaria:2007bb, Beccaria:2009eq, Lukowski:2009ce} for more details, the number of terms to be fitted against and the complexity of the result still increase rapidly with the loop order. This method has allowed for the construction of the anomalous dimension for any $M$ for twist two and three operators up to $5$ loops, we refer to the mentioned papers for explicit expressions for the anomalous dimensions as the expressions grow rapidly in length and fill pages. There are also some results for operators with higher twist \cite{Beccaria:2008pp}.

Another complementary approach has been to study the so called Baxter equation. This equation is formulated in terms of the Baxter function, $Q(u)=\prod_{j=1}^M(u-u_j)$, and the transfer matrix eigenvalue, $t(u)$, and reads for the $\mathfrak{sl}(2)$ sector at one loop
\begin{eqnarray}
&&t(u)Q(u)=(u+i/2)^LQ(u+i)+(u-i/2)^LQ(u-i),\\
&&t(u)=2u^L+q_{L-2}u^{L-2}+\dots+q_0, 
\end{eqnarray}
where $q_{L-2}=-(M+L/2)(M+L/2+1)-L/4$. The remaining $q_r$, $r=0,\dots,L-3$ specify the state. For $L=2$ there is only one state and this equation can be identified with the equation for Wilson or Hahn polynomials and is hence solved by \cite{Korchemsky:1995be, Eden:2006rx}
\begin{equation}
Q(u)={}_4F_3\left(-\tfrac{M}{2},\tfrac{M+1}{2},\tfrac{1}{2}+iu,\tfrac{1}{2}-iu;1,1,\tfrac{1}{2};1\right)={}_3F_2\left(-M,M+1,\tfrac{1}{2}-iu;1,1;1\right).
\end{equation}
The one-loop anomalous dimension is given by
\begin{equation}
\gamma(g,M)=\left.2g^2\frac{d}{du}\left(i\log Q(u+i/2)\right)\right|_{u=0}=8g^2 S_1(M)
\end{equation}
which at large $M$ reduces to \eqref{cuspAD}. The analogous result for the ground state with $L=3$ was found in \cite{Derkachov:1999ze}.

The all loop $\mathfrak{sl}(2)$ Baxter equation was derived in a series of papers \cite{Belitsky:2006av,Belitsky:2006wg,Belitsky:2009mu} and this construction made it possible to develop methods for finding higher loop solutions exact in the spin $M$. By a deformation of the one-loop solution the result for $L=2,3$ was found to three loops \cite{Kotikov:2008pv} and later extended to four loops \cite{Beccaria:2009rw}.

The construction of these results exact in spin allowed for another important check of the Bethe equations. The asymptotic equations are supposed to break down due to wrapping effects. As the interaction range is growing with the order in the loop expansion this happens soon enough for short operators as twist two and three. Nevertheless there are indications that terms up to the order $\mathcal{O}((\tfrac{\log M}{M})^2)$ in the large spin expansion are free from corrections. To investigate the supposed breakdown of the Bethe ansatz prediction at higher loop orders the studies of high energy scattering amplitudes in $\mathcal{N}=4$ provides important insights. The Balitsky-Fadin-Kuraev-Lipatov (BFKL) equation describes the high energy scattering both in QCD and in $\mathcal{N}=4$ and provides the relation between the anomalous dimension and the spin $M$ near the point $M=-1$ \cite{Kuraev:1977fs,Balitsky:1978ic,Kotikov:2002ab}. With the spin $M=-1+\omega$, and $\omega$ taken small, the one-loop relation for twist 2 operators reads
\begin{equation}
-\frac{\omega}{4g^2}=\psi\left(-\tfrac{\gamma}{2}\right)+\psi\left(1+\tfrac{\gamma}{2}\right)-2\psi(1),
\end{equation}
Expanding the $\psi$-functions in power series and then inverting the series we find 
\begin{equation}
\gamma=2\left(-\frac{4g^2}{\omega}\right)-4\zeta(3)\left(-\frac{4g^2}{\omega}\right)^4+\mathcal{O}\left(g^{12}\right).
\end{equation}
Hence the one-loop BFKL equation provides an all-loop prediction for the leading singularities as $\omega\to 0$. The 5-loop result for general values of the spin can now be analytically continued to $M=-1+\omega$ and expanded for $\omega\ll 1$, the result is
\begin{equation}
\gamma=2\left(-\frac{4g^2}{\omega}\right)-2\frac{\left(-4g^2\right)^4}{\omega^7}+2\frac{\left(-4g^2\right)^5}{\omega^9}+\mathcal{O}\left(g^{12}\right).
\end{equation}
This clearly shows that the Bethe ansatz prediction breaks down at four loops, as was indeed expected from general considerations.  It is also possible to check that the results are free from wrapping up to $\mathcal{O}((\tfrac{\log M}{M})^2)$, for $L=2$ to four loops this has been demonstrated explicitly \cite{Bajnok:2008qj,Beccaria:2009vt} and for 5-loops this is very likely to hold \cite{Lukowski:2009ce}. Similar arguments can also be made for $L=3$ \cite{Beccaria:2009eq}.
 Furthermore the BFKL equation has been proposed to two-loop order which by the same arguments as above provides an all-loop prediction for the next to leading singularity, also this prediction disagrees with the result from the Bethe ansatz \cite{Kotikov:2007cy, Lukowski:2009ce}.
Including wrapping corrections restores this agreement, see \cite{chapLuescher} for a review.

The exact expressions in terms of harmonic sums as well as the integral equations for further subleading terms in the large $M$ expansion show yet another interesting feature.  Continuing the expansion beyond the leading order one finds that the perturbative expressions can be organised as in \eqref{twist2subleading} \footnote{A similar structure is also present for $L>2$ and for the further terms in the large $M$ expansion.}. The structure observed in the expansion is due to a property referred to as reciprocity or parity preservation \cite{Dokshitzer:2005bf,Basso:2006nk,Dokshitzer:2006nm}, see also \cite{Beccaria:2010tb} for a recent review.
The twist operators \eqref{twistop} 
 can be classified using representations of the collinear $SL(2,\mathbb{R})$ subgroup of the conformal group $SO(2,4)$ \cite{Braun:2003rp, chapQCD}. This suggests the conformal spin, $m=M+L/2+\gamma(M)$, as a natural parameter for the expansion and that the anomalous dimension can be written as
\begin{equation}
\gamma(M,L)=f(M+\gamma(M,L),L).
\end{equation}
Reciprocity states that the function $f(M)$ can be expanded in terms of the quadratic Casimir of the collinear group, $J^2=(M+L/2)(M+L/2-1)$
\begin{equation}\label{reciprocity} 
f(M)=\sum_{n=0}^\infty \frac{f_n(\log J)}{J^{2n}}.
\end{equation}
This is fully consistent with the relations between the coefficients in \eqref{twist2subleading} and implies relations between further terms in the expansion and hence reduces the number of unknown functions appearing in the expansion, roughly by a factor of two.

Reciprocity also restricts the expressions exact in $M$ which naturally is helpful for finding them, it can be used to restrict the ansatz in terms of harmonic sums \cite{Beccaria:2008fi,Beccaria:2009eq,Beccaria:2009rw,Lukowski:2009ce}. An interesting thing to note however is that the perturbative expressions found respects reciprocity both when wrapping interactions are included as well as when they are not \cite{Beccaria:2009vt}. 

Reciprocity in this context is based on an earlier idea that appeared in the context of deep inelastic scattering in QCD, the Gribov-Lipatov one-loop reciprocity \cite{Gribov:1972rt}. That idea states that the splitting functions $P(x)$, related to the twist 2 anomalous dimensions in QCD by Mellin transformation, obeys the relation $P(x)=-xP(1/x)$. This leads to relations between coefficients in the large $M$ expansion of the anomalous dimension as were first observed in \cite{Moch:2004pa,Vogt:2004mw}, the so called MVV relations, and extended to higher orders in the expansion in  \cite{Basso:2006nk}. 

A complete understanding of the origin of the observed reciprocity property is lacking. It is not a unique feature of $\mathcal{N}=4$ SYM, not tied to integrability or even to the planar limit. Indeed it holds in QCD in sectors where integrability is not present and for an arbitrary number of colors \cite{Basso:2006nk}. In $\mathcal{N}=4$ reciprocity is a feature of other known minimal anomalous dimensions of twist operators \cite{Beccaria:2007vh,Beccaria:2007bb,Beccaria:2007pb,Beccaria:2008fi} while broken for higher states in the band \cite{Belitsky:2008mg,Giombi:2009gd}. Further it was also observed in $\mathcal{N}=6$ Chern-Simons theory \cite{Giombi:2009gd}, see also the review \cite{chapN6}.

\section{Strong coupling expansion}

With the equation \eqref{BES} we have the first explicit realisation of a quantity important for the AdS/CFT correspondence that can be computed for all values of the coupling constant. Considering the dual string states we can therefore put both the Bethe equations and the correspondence itself to a test. Studying the dual string states, folded strings with one large angular momentum $S$ on $AdS_3$, the string energy could be computed to first orders in the strong coupling expansion \cite{Gubser:2002tv,Frolov:2002av,Kruczenski:2002fb,Roiban:2007jf,Roiban:2007dq,Giombi:2009gd, Beccaria:2010ry} (see also \cite{chapSpinning,chapQString}), this provides the prediction for the scaling function
\begin{equation}\label{strongcoupling}
f(g)=4g-\frac{3\log 2}{\pi}-\frac{K}{4\pi^2}\frac{1}{g}+\dots,
\end{equation}
where $K$ is Catalan's constant.
It is thus desirable to solve \eqref{BES} at strong coupling. This however turned out to be a lot harder than first expected. The integral equation was shown to reproduce the first two orders in the strong coupling expansion numerically with high accuracy \cite{Benna:2006nd}. Despite this a straightforward expansion gives only the leading term in \eqref{strongcoupling} analytically, already the first subleading term could not be obtained analytically this way. With considerable effort the first subleading term was reproduced by solving the Bethe equations \cite{Casteill:2007ct,Belitsky:2007kf}, however this was done by first expanding the equations at strong coupling and then taking the large spin limit and hence not by solving the integral equation \eqref{BES}. This approach can not easily be continued to include higher orders in the expansion. Some important progress towards solving the integral equation, analytical as well as numerical, was made in \cite{Alday:2007qf,Kostov:2007kx,Kotikov:2008zz,Beccaria:2007tk} and the final solution was then presented in \cite{Basso:2007wd} and further discussed in \cite{Kostov:2008ax,Basso:2009gh}.

In order to solve \eqref{BES} at strong coupling it proved useful to split the density as
\begin{equation}\label{split}
\frac{e^t-1}{t}\hat{\sigma}(t)=\frac{\gamma_+(2gt)}{2gt}+\frac{\gamma_-(2gt)}{2gt}
\end{equation}
and expand the even and odd part in Neumann series \cite{Alday:2007qf, Basso:2007wd}
\begin{eqnarray}\label{Neumanndec}
&&\nonumber\gamma_+(t)=\sum_{k=1}^\infty (-1)^{k+1}2k\,J_{2k}(t)\gamma_{2k}\\
&&\gamma_-(t)=\sum_{k=1}^\infty (-1)^{k+1}(2k-1)J_{2k-1}(t)\gamma_{2k-1}.
\end{eqnarray}
Applying the density split \eqref{split} in \eqref{BES} the equation can be decomposed into an even and an odd part \cite{Kotikov:2008zz,Eden,Basso:2007wd}.
Introducing a further change of variables,
\begin{eqnarray}\label{BKKvariables}
\Gamma(t)=\left(1+i\coth\frac{t}{4g}\right)\gamma(t)
\end{eqnarray}
where $\Gamma(t)=\Gamma_+(t)+i\Gamma_-(t)$ and $\gamma(t)=\gamma_+(t)+i\gamma_-(t)$, it is possible to collect all dependence on the coupling constant in $\Gamma_\pm(t)$ and \eqref{BES} is rewritten as the system
\begin{eqnarray}\label{BKK}
&&\nonumber \int_0^\infty\frac{dt}{t}\left(\Gamma_+(t)+\Gamma_-(t)\right)J_{2n}(t)=0\\
&&\int_0^\infty\frac{dt}{t}\left(\Gamma_-(t)-\Gamma_+(t)\right)J_{2n-1}(t)=\delta_{n,1}.
\end{eqnarray}
The coupling constant dependence of the functions $\Gamma_\pm(t)$ is determined by requiring the correct analyticity properties of the functions, which follows from \eqref{BKKvariables} and \eqref{Neumanndec}. \eqref{BKKvariables} can be rewritten and expanded as \cite{Basso:2009gh}
\begin{equation}\label{BKKvariables2}
\Gamma(it)=\frac{\sin(\tfrac{t}{4g}+\tfrac{\pi}{4})}{\sin(\tfrac{t}{4g})\sin(\tfrac{\pi}{4})}\gamma(it)=\gamma(it)\sqrt{2}\prod_{k=-\infty}^\infty\frac{t-4\pi g(k-1/4)}{t-4\pi g k}.
\end{equation} 
From this one can conclude \cite{Kostov:2008ax,Basso:2009gh} that $\Gamma(it)$ has an infinite set of zeros and poles given by
\begin{eqnarray}
&&t_{zero}=4\pi g(l-1/4), \quad l\in \mathbb{Z}\label{zeros}\\
&&t_{pole}=4\pi g l', \quad l'\in \mathbb{Z},\,l'\neq0.
\end{eqnarray}
To construct the general solution to the integral equations we will consider the inverse Fourier transform of $\Gamma(t)$, $\Gamma(u)=\int_{-\infty}^\infty\tfrac{dt}{2\pi}e^{iut}\Gamma(t)$. From \eqref{BKKvariables2} follows
\begin{equation}
\Gamma(u)=\int_{-\infty}^\infty\frac{dt}{2\pi}\,e^{iut}\,\frac{\sinh(\tfrac{t}{4g}+i\tfrac{\pi}{4})}{\sin(\tfrac{t}{4g})\sin(\tfrac{\pi}{4})}\gamma(t).
\end{equation}
This integral can be computed by deforming the contour, picking up the residues of the poles along the imaginary axis.
This can however only be done if the integrand vanishes at infinity. Since $\gamma(t)$ admits a Neumann expansion we can conclude that $\gamma(u)=0$ for $u^2>1$, using the property of the Bessel functions that $\int_{-\infty}^\infty \tfrac{dt}{2\pi}e^{iut}J_n(t)=0$ for $u^2>1$. This means that $\gamma(t)\sim e^{|t|}$ for large complex $t$ as $|t|\to\infty$ and hence the above integral can be computed by deforming the contour when $u^2>1$, the result is
\begin{equation}\label{Gammasol1}
\Gamma(u)=\theta(u-1)\sum_{n=1}^\infty c_+(n,g)e^{-4\pi n g(u-1)}+\theta(-u-1)\sum_{n=1}^\infty c_-(n,g)e^{-4\pi n g(-u-1)}
\end{equation}
where $c_\pm(n,g)=\mp4g\gamma(\pm 4\pi i gn)e^{-4\pi n g}$.
To find $\Gamma(u)$ for $-1\leq u\leq1$ we will make an attempt to solve the equations \eqref{BKK}. Before doing that we note that the infinite system in \eqref{BKK} can be rewritten as two equations with an additional arbitrary parameter $\phi$ by applying the relation derived from the Jacobi-Anger identity,
\begin{equation}
e^{it\sin\phi}=\frac{2}{\cos\phi}\sum_{n=1}^\infty\left(\tfrac{J_{2n-1}(t)}{t}(2n-1)\cos((2n-1)\phi)+i\tfrac{J_{2n}(t)}{t}2n\sin(2n\phi)\right).
\end{equation}
Using the notation $u=\sin\phi$ we find for $-1\leq u\leq1$
\begin{eqnarray}
\int_0^\infty dt\left(e^{itu}\Gamma_-(t)-e^{-itu}\Gamma_+(t)\right)=2.
\end{eqnarray}
In terms of $\Gamma(u)$ this equation can be expressed as
\begin{eqnarray}
&&\Gamma(u)+\frac{1}{\pi}\pint_{-1}^1dv\frac{\Gamma(v)}{v-u}=\Phi(u)\\
&&\Phi(u)=-\frac{1}{\pi}\left(2+\int_{-\infty}^{-1}dv\frac{\Gamma(v)}{v-u}+\int_1^\infty dv\frac{\Gamma(v)}{v-u}\right).
\end{eqnarray}
In $\Phi(u)$ one can use the solution \eqref{Gammasol1} and solving the above equation renders a solution in terms of $c_\pm(n,g)$, valid in the interval $-1\leq u\leq1$. Since this solution together with \eqref{Gammasol1} determines the function $\Gamma(u)$ for all values of $u$ it is now finally possible to go back to the Fourier transform and find an expression for $\Gamma(it)$, we refer to \cite{Basso:2009gh} for details.

The general solution is expressed in terms of the functions of the coupling constant $c_\pm(n,g)$. These are completely determined by requiring the correct analyticity properties of the function $\Gamma(it)$, requiring that the function has zeros according to \eqref{zeros} we find the so called quantisation conditions.

These conditions determine $c_\pm(n,g)$ and the solution for all values of the coupling constant. Expanding the quantization condition at strong coupling one finds, to leading orders,
\begin{eqnarray}
c_+(n,g)&=&(8\pi g n)^{1/4}\frac{2\Gamma(n+1/4)}{\Gamma(n+1)\Gamma^2(1/4)}\left(1-\frac{1}{g}\left(\frac{3\log 2}{4}+\frac{3}{32n}+\dots\right)\right)\\
c_-(n,g)&=&(8\pi g n)^{1/4}\frac{\Gamma(n+3/4)}{2\Gamma(n+1)\Gamma^2(3/4)}\left(1+\frac{1}{g}\left(\frac{3\log 2}{4}+\frac{5}{32n}+\dots\right)\right).
\end{eqnarray}
Finally the scaling function is written in terms of $c_\pm(n,g)$ and from this the strong coupling expansion \eqref{strongcoupling} follows.

It is also possible to continue this expansion to many more orders in the coupling constant \cite{Basso:2007wd} (see also \cite{Volin:2010cq} for a Mathematica code that generates the expansion). The first orders above are the ones that have been checked against the semiclassical quantisation of the string sigma model, comparison of further terms remains a challenge.

Solving the integral equation for subleading corrections in the large spin expansion at strong coupling turns out to be straightforward once the analysis for the leading term is completed. It is possible to write the first subleading terms in terms of the leading solution. The coefficients $c_\pm(n,g)$ determine the anomalous dimension to order $\mathcal{O}(\frac{1}{(\log M)^k})$, $k\to\infty$ for general values of $L$ and up to order $\mathcal{O}((\tfrac{\log M}{M})^2)$ for $L=2,3$.  It is interesting to note that when evaluating the subleading terms at strong coupling one finds that the expansion reorganises in terms of the parameter $M/g$ \cite{Freyhult:2009my}. This is fully consistent with the computation done in string theory where $S/g$ is kept fixed expanding for large $g$ and then subsequently taken large. For twist two the string result \cite{Beccaria:2008tg} is reproduced from the Bethe ansatz, since the computation is done by resumming all orders in the weak coupling expansion this gives a strong indication that wrapping plays no role for these first orders in the large $M$ expansion.

As for the weak coupling expansion reciprocity is suggested to hold for the all loop anomalous dimension. The energy of the dual string solutions exhibits the same structure as the anomalous dimensions \eqref{twist2subleading} and reciprocity is verified at strong coupling \cite{Basso:2006nk,Beccaria:2008tg,Beccaria:2010ry}. Indeed this is also confirmed by the structure obtained by solving the integral equations that follow from the asymptotic Bethe equations up to $\mathcal{O}((\frac{\log M}{M})^2)$ \cite{Freyhult:2009my}. 

Thanks to the method developed to treat the strong coupling expansion of the BES equation also other closely related integral equations, corresponding to operators belonging to other larger sectors of the theory, could be solved \cite{Basso:2010in,Freyhult:2009fc} and the anomalous dimensions matched to the corresponding string states \cite{chapSpinning}. This was possible because the kernels coincided with the BES kernel. In sectors where that is not the case the strong coupling solutions of integral equations derived from the Bethe equations remain a challenge \cite{Rej:2007vm}.

\section{Non-perturbative corrections and intermediate values of the coupling constant}
The quantization condition as well as the expression for the scaling function in terms of the functions $c_\pm(n,g)$ are valid for all values of the coupling constant. The equations do not admit a solution on a closed form for arbitrary coupling but numerically it is however possible to find a solution for any value of the coupling constant which indeed demonstrates a smoothly interpolating scaling function \cite{Basso:2007wd,Kotanski:2008tv,Basso:2009gh}.
The strong coupling expansion in \eqref{strongcoupling} defines an symptotic series and is non-Borel summable. The reason for this is that when expanding the quantization conditions at strong coupling we have not taken non-perturbative corrections into account. 
Using the construction in the previous section and expanding the quantisation conditions, retaining the non-perturbative corrections, it is found to leading orders \cite{Basso:2009gh}
\begin{equation}\label{nonperturbative}
f(g)=4g-\frac{3\log 2}{\pi}-\frac{K}{4\pi^2 g}+\mathcal{O}(1/g^2)-\frac{2\Lambda^2}{\pi^2}\left(1+\frac{3-6\log 2}{16\pi g}+\mathcal{O}(1/g^2)\right)+\mathcal{O}(\Lambda^4)
\end{equation}
where
\begin{equation}
\Lambda^2=\sigma\frac{1}{\sqrt{2\pi g}}e^{-2\pi g}\frac{\Gamma(\tfrac{3}{4})}{\Gamma(\tfrac{5}{4})}
\end{equation}
defines the non-perturbative scale. The non-perturbative corrections can be organised in terms of the mass of the $O(6)$-model
\begin{equation}
m_{O(6)}=\frac{4\sqrt 2}{\pi\sigma}\Lambda^2\left(1+\frac{3-6\log 2}{16\pi g}+\mathcal{O}(1/g^2)\right)+\mathcal{O}(\Lambda^4).
\end{equation}
Note that the series in $1/g$ in \eqref{nonperturbative} is not Borel summable. It is only the series in $1/g$ together with the expansion in $\Lambda^2$ that is Borel summable. The complex parameter $\sigma$ appearing in the expansion is determined by the way the series in $1/g$ is regularised, alternatively on the renormalisation scheme used in the $O(6)$-model. 

The cusp anomalous dimension appears also in the leading order of the large $M$ expansion of twist operators with $L=j \log M$, where $j$ is some finite number. The leading order is governed by the the so called generalised scaling function of which the scaling function constitutes a part \cite{Belitsky:2006en,Alday:2007mf,Freyhult:2007pz,Basso:2008tx,Bajnok:2008it},
\begin{equation}
\gamma(g,j,M)=\left(f(g)+\epsilon(g,j)\right)\log M.
\end{equation}
The function $\epsilon(g,j)$ denotes the twist dependent part of the generalised scaling function.
The string duals are folded strings spinning with one angular momentum, $S$ identified with $M$, in $AdS_5$ and one angular momentum, $J\propto \log S$ identified with $L$, on $S^5$. The energy of the string solutions was obtained directly from the string sigma model to two-loops in the sigma model loop expansion, see \cite{Giombi:2010fa} and references therein.
The string sigma model with $J\sim \log S$, $S$ and $g$ taken large reduces to the $O(6)$ model \cite{Alday:2007mf}. This corresponds to a low energy limit in which only the massless excitations around the classical solution remain. These massless fields describe the $O(6)$ model and as a result the sigma model can be completely solved in this limit. The limit can also be considered from the gauge theory side using the asymptotic Bethe equations \cite{Freyhult:2007pz} where the relation to the sigma model was explicitly demonstrated in \cite{Basso:2008tx} and further studied in \cite{Fioravanti:2008ak,Buccheri:2008ap,Gromov:2008en,Beccaria:2008nf,Fioravanti:2008bh,Bajnok:2008it,Volin:2008kd}. The generalised scaling function is hence associated with the free energy of the $O(6)$-sigma model, a quantity that receives non-perturbative corrections in terms of the mass scale of the sigma model $m_{O(6)}$.
  
In \cite{Basso:2009gh} it was demonstrated that the non-perturbative scale appearing in the strong coupling expansion of the scaling function is related to the non-perturbative scale, the mass gap, in the $O(6)$-model. In fact the mass scale that can be identified in $f(g)$ is identical to $m_{O(6)}$ to all orders in the strong coupling expansion.

\phantomsection
\addcontentsline{toc}{section}{\refname}


\end{document}